\documentclass[aps,prl,reprint,amsmath,amssymb,showpacs,superscriptaddress,twocolumn]{revtex4-2}
\usepackage{graphicx}
\usepackage{xcolor}
\usepackage{subfigure}
\usepackage{hyperref,hypcap}
\usepackage{braket}
\usepackage{amsmath}
\usepackage{commath}
\usepackage{mathtools}

\newcommand{\rYlm}{{\mathcal Y}_{\ell}^M}

\begin{document}

\title{Imaging the valley and orbital Hall effect in monolayer MoS$_2$}

	\author{Fei Xue} 
	\affiliation{Physical Measurement Laboratory, National Institute of Standards and Technology, Gaithersburg, MD 20899, USA}
	\affiliation{Institute for Research in Electronics and Applied Physics \& Maryland Nanocenter,	University of Maryland, College Park, MD 20742}
	\author{Vivek Amin}	
	\affiliation{Physical Measurement Laboratory, National Institute of Standards and Technology, Gaithersburg, MD 20899, USA}
	\affiliation{Institute for Research in Electronics and Applied Physics \& Maryland Nanocenter,	University of Maryland, College Park, MD 20742}
	\author{Paul M. Haney} 
	\affiliation{Physical Measurement Laboratory, National Institute of Standards and Technology, Gaithersburg, MD 20899, USA}

	\date{\today}
	
	\begin{abstract}

The topological properties of a material's electronic structure are encoded in its Berry curvature, a quantity which is intimately related to the transverse electrical conductivity. In transition metal dichalcogenides with broken inversion symmetry, the nonzero Berry curvature results in a valley Hall effect.  In this paper we identify a previously unrecognized consequence of Berry curvature in these materials: an electric field-induced change in the electrons' charge density orientation.  We use first principles calculations to show that measurements of the electric field-induced change in the charge density or local density of states in MoS$_2$ can be used to measure its energy-dependent valley and orbital Hall conductivity.

	\end{abstract}
	
\maketitle

{\it Introduction} --  Transition metal dichalcogenides (TMDs) exhibit a wide array of novel phenomena related to their valley degree of freedom \cite{xiao2012coupled}.  The valley index labels one of multiple degenerate energy extrema in the conduction or valence band.  Phases of TMDs which lack inversion symmetry possess nontrivial Berry curvature - a quantity that determines the transverse electrical conductivity and which describes the topological properties of the electronic structure \cite{RMP_Berry}.  States in different valleys have opposite sign of Berry curvature and therefore flow in opposite directions transverse to an applied electric field. This results in vanishing net transverse charge current (or Hall current), but a nonzero valley Hall current \cite{xiao2012coupled}.  The valley Hall effect is accompanied by accumulation of orbital angular momentum at sample boundaries, which has been measured with the magneto-optical Kerr effect \cite{lee2016electrical}. Other experimental probes of the valley Hall effect use an applied magnetic field \cite{srivastava2015valley} or circularly polarized light to generate a valley-polarized state \cite{ubrig2017microscopic,mak2012control,mak2014valley}, enabling a nonzero transverse Hall current.  The valley Hall effect has also been detected with nonlocal voltage measurements \cite{hung2019direct,wu2019intrinsic}.

In this work we introduce a new observable associated with the valley Hall effect in TMD's: an electric field-induced change in the bulk charge density distribution. More generally, we demonstrate that in a specified class of lattices, the electric field-induced change in a state's charge density distribution is directly related to its Berry curvature. Using first principles calculations of MoS$_2$ as a representative example, we show that measurements of the electric field-induced change in density provide a quantitative estimate of the valley Hall conductivity. We also propose that spectroscopic measurements of the electric field-induced change in the local density of states enables a measurement of the energy-resolved valley Hall conductivity.\\

{\it Formalism} -- We first present the general relation between a state's Berry curvature and the electric field-induced change in its real space density.  This analysis applies to lattices composed of a monoatomic unit cell with orbitals of the same character ({\it e.g.}, $p$-like or $d$-like).  Although these are restrictive assumptions, they pertain to cases of practical interest, such as near-band edge states of TMDs in the 2H phase \cite{xiao2012coupled}.  We begin with the relation between a state's real space density $\rho({\bf r})$ and its velocity.  The periodic lattice Hamiltonian is represented in a basis of real localized orbitals, such as Wannier functions or atomic orbitals, which are
a product of a radial function and angular function: $\phi_M({\bf r}) = R_{n,\ell}\left(r\right)\rYlm\left(\theta,\phi\right)$, where $R_{n,\ell}(r)$ is a radial function for states in the $n^{\rm th}$ shell with angular momentum $\ell$.  $\rYlm$ is the real-valued spherical harmonic, with $2\ell +1$ distinct $M$ labels ({\it e.g.}, for $\ell=1$, $M=\{x,y,z\}$).  Note that $M$ is sufficient to label the basis states since we assume all orbitals have the same value for $n$ and $\ell$.  In this basis, the tight-binding crystal field Hamiltonian is:
\begin{equation}
H_0({\bf k})=\sum_{{\bf R},M,M'}\cos({\bf k}\cdot{\bf R})~ t_{M,M'}^\sigma({\bf R}) \left(c_{M}^{\dagger}c_{M'}+h.c.\right),
\label{eq:H0}
\end{equation}
where ${\bf k}$ is the crystal momentum, and the sum ${\bf R}$ includes the primary unit cell (${\bf R}=0$) and all other Bravais lattice vectors. Typically the sum is truncated up to the n$^{\rm th}$ nearest-neighbor. $c_{M}^{\dagger}(c_{M'})$ is the electron creation (annihilation) operator for orbital $M$($M'$) located in the primary unit cell.  $t_{M,M'}^\sigma\left({\bf R}\right)$ is the $\sigma$-hopping parameter between orbital $M$ of the primary unit cell and orbital $M'$ centered at lattice site ${\bf R}$.    Eq.~\ref{eq:H0} includes only $\sigma$-hopping, however generally $\pi$ and $\delta$ hopping amplitudes are smaller, so that the above model is typically adequate (see Supp. Info for a comprehensive assessment of the model's validity).

The velocity operator ${\bf v}=dH/d{\bf k}$ derived from this Hamiltonian is:
\begin{equation}
{\bf v}({\bf k})=-\sum_{{\bf R},M,M'}{\bf R}\sin({\bf k}\cdot{\bf R})t_{M,M'}^\sigma({\bf R}) \left(c_{M}^{\dagger}c_{M'}+h.c.\right).
\label{eq:v}
\end{equation}
Each term in the sum over ${\bf R}$ represents the current flowing between the atom at ${\bf R}=0$ and the atom at ${\bf R}$.  Note that a current between sites separated by ${\bf R}$ requires nonzero overlap between ${\bf k}$ and ${\bf R}$, and overlap between orbitals. 
The orbital overlap is closely related to the shape of the wave function density $\rho_0({\bf r})$:
\begin{equation}
\rho_0({\bf r}) = \sum_{M,M'} \left(c_{M}^{\dagger}c_{M'}+h.c.\right) ~\phi_M\left({\bf r}\right)
    \phi_{M'}\left({\bf r}\right).
\label{eq:rho}
\end{equation}
$\rho_0$ corresponds to the density contribution from ``on-site'' orbitals centered at ${\bf R}=0$, and omits contributions from overlap between neighboring sites in the lattice. In the Supp. Info we show that the differences between the total and ``on-site'' densities are small in the regions of interest for this work.  The form of the orbital basis implies that $\rho_0$ is the product of a radial function and an angular function: $\rho_0({\bf r}) =\rho_0^{\rm rad}(r)~\rho_0^{\rm ang}({\bf {\hat r}})$, where the atom center is at $r=0$, and $\rho_0^{\rm rad}(r)$ is normalized such that $\int_0^\infty r^2 \rho^{\rm rad}(r)dr=1$ (so that $\rho_0^{\rm ang}({\bf {\hat r}})$ is dimensionless).

In the Supp. Info, we show that the angular part of the density $\rho_0^{\rm ang}({\bf {\hat r}})$ is related to the hopping parameters and wave function coefficients of Eq. \ref{eq:v} (the last two factors in that equation).  This leads to the following relation between a state's velocity and density:
\begin{eqnarray}
{\bf v}({\bf k})= & -\sum_{{\bf R}}{\bf R}\sin({\bf k}\cdot{\bf R})~\tilde{t}^\sigma(R) \rho_0^{\rm ang}({\bf k},{\bf {\hat R}}).\label{eq:vrho}
\end{eqnarray}
where $\tilde{t}^\sigma(R)={t}^\sigma(R)/Y_\ell^{m=0}(0,0)^2$ is the $\sigma$-hopping parameter normalized by a (known) constant factor.  Note that we added a ${\bf k}$ label for $\rho$; we explicitly include this argument for all ${\bf k}$-dependent quantities in the paper. $t^\sigma(R)$ is the $R$-dependent $\sigma$-hopping function whose form is semi-universal and taken as known {\it a priori}~\cite{harrison2012electronic}. In the sum over neighboring atom positions ${\bf R}$, a positive current corresponds to a state propagating outward from ${\bf r=0}$ to ${\bf r}={\bf R}$, while a negative current corresponds to a state propagating inward from ${\bf r}={\bf R}$ to ${\bf r}=0$.

Eq.~\ref{eq:vrho} is one of our primary results, and provides an intuitive relationship between a state's charge density and current: the velocity along ${\bf {\hat R}}$ is the product of the wave function phase change along ${\bf {\hat R}}$ (given by the factor $\sin({\bf k} \cdot {\bf R})$) and the density along ${\bf {\hat R}}$ (given by the factor $\rho_0^{\rm ang}({\bf k}, {\hat{ \bf R}})$).   This representation of velocity is a substantial simplification of the general form given in Eq. \ref{eq:v}, which requires knowledge of the full hopping elements (encoded in Slater-Koster tight-binding tables\cite{SlaterKoster}) and the complex wave function coefficients.

Armed with this density-velocity connection, we next turn to the relation between a state's Berry curvature and its velocity.  The $z$-component of the Berry curvature $\Omega_z^n$ of an eigenstate $\psi_n$ is \cite{nagaosa2010anomalous}:
\begin{equation}
	\Omega_n^z({\bf k})=-2 ~\text{Im}\sum_{m \ne n}\frac{\braket{\psi_{n{\bf k}}|v_x|\psi_{m{\bf k}}}\braket{\psi_{m{\bf k}}|v_y|\psi_{n{\bf k}}}}{(\epsilon_m-\epsilon_n)^2},
	\label{eq:berry2}
\end{equation}
where $\epsilon_n$ is the energy of the $n^{\rm th}$ eigenstate $|\psi_{n{\bf k}}\rangle$, and $v_{x,y}=dH/dk_{x,y}$.  The intrinsic contribution to the anomalous Hall conductivity $\sigma_{\rm AHE}$ in the clean limit is given by the sum of the occupied states' Berry curvature.  In this work, we also consider the valley Hall and orbital Hall effects.  The expression for the valley Hall conductivity $\sigma_{\rm OHE}$ is the same as Eq. \ref{eq:berry2} with an additional factor of $\pm 1$ depending on the state's valley index ({\it e.g.}, the location of the ${\bf k}$-point in the Brillouin zone, see Sec. II).  We also compute the orbital Hall conductivity, which describes the electric field-induced flow of orbital angular momentum, oriented in the $z$-direction and with velocity transverse to the applied field \cite{bernevig2005orbitronics,go2018intrinsic,bhowal2020intrinsic,canonico2020orbital}. The orbital Hall conductivity is also given by a Kubo formula expression like Eq. \ref{eq:berry2}, with the replacement $v_y\rightarrow \left(v_y L_z + L_z v_y\right)/2$, where $L_z$ is the atomic orbital angular momentum operator. Finally, we note that the Berry curvature may be nonzero only in the presence of time reversal and/or inversion symmetry breaking \cite{xiao2007valley}.

A state's Berry curvature can also be understood in terms of perturbation theory.  An applied electric field perturbs the eigenstates and may change their velocity.  The linear-in-$E$ change in velocity $\delta {\bf v}$ determines the Berry curvature: ${\bf \Omega}_n({\bf k}) = {\hat {\bf E}}\times\langle \delta {\bf v}_n({\bf k}) \rangle$. We invoke Eq.~\ref{eq:vrho} to relate the change in velocity to a change in the charge density.  Using this expression for the Berry curvature, we conclude that the real space density response of the n$^{\rm th}$ state is related to its Berry curvature:
\begin{equation}
{\bf \Omega}_n({\bf k}) = -{\hat {\bf E}}\times\sum_{{\bf R}}{\bf R}\sin({\bf k}\cdot{\bf R})~\tilde{t}^\sigma(R) ~\langle \delta \rho_{0,n}^{\rm ang}({\bf k},{\hat {\bf R}}) \rangle.
\label{eq:berry3}
\end{equation}
Eq.~\ref{eq:berry3} is another primary result, which provides the connection between a state's Berry curvature and the E-field-induced change in its density distribution.  We emphasize that this relation is valid for lattices with monoatomic unit cells and orbitals with the same character.

It's straightforward to show that the net changes in velocity and density are derived from mutually exclusive sets of states.  The net change of any observable is obtained by summing over ${\bf k}$.  The $\sin({\bf k}\cdot {\bf R})$ factor on the right hand side of Eq. \ref{eq:vrho} indicates that ${\bf v}({\bf k})$ and $\delta \rho_0({\bf k},{\bf R})$ have opposite parity under ${\bf k} \rightarrow -{\bf k}$.  The anomalous Hall conductivity is derived from states with even-in-${\bf k}$ $\Omega({\bf k})$.  However these states' $\delta\rho_0({\bf k})$ is odd-in-${\bf k}$, and therefore make no contribution to $\delta\rho^{\rm tot}_0$.  The converse also holds: states which contribute to $\delta\rho^{\rm tot}_0$ do not contribute to $\sigma_{\rm AHE}$.

We finally note recent work which formulates the anomalous Hall conductivity as a local property without reference to Bloch wave vector ${\bf k}$ \cite{marrazzo2017locality,bianco2011mapping,caio2019topological}.  The formal structure of this theory includes the response of the second order cumulant of the charge density, resembling the picture we describe of a change in the charge density distribution.  The present work straddles between the limiting cases of the more standard formulation of Berry curvature strictly in ${\bf k}$-space and the work \cite{marrazzo2017locality,bianco2011mapping} which resides entirely in real space.\\

\begin{figure}
    \includegraphics[width=.9\columnwidth]{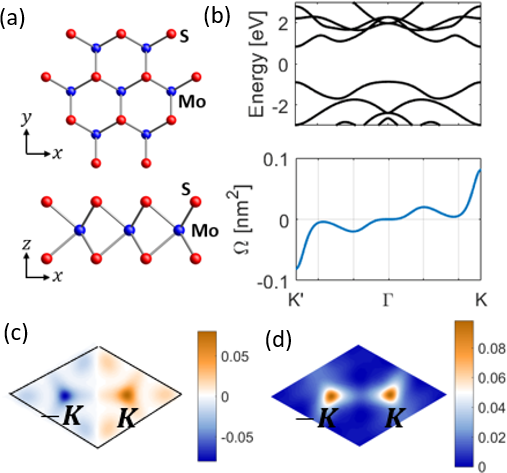}
	\caption{(a) depicts the 2H-MoS$_2$ monoloayer. (b) shows the band structure and Berry curvature along high symmetry lines. (c) and (d) show the Berry curvature and the orbital Hall conductivity in the Brillouin zone (units of ${\rm nm}^2$).} \label{fig:TMDbands}
\end{figure}

{\it Imaging the valley Hall effect in insulating TMD} -- As an application of our formalism, we consider a monolayer of MoS$_2$ in the 2H phase.  We'll show the relation between the electric field-induced change in density and the valley Hall effect for three cases of increasing complexity: the response at a single ${\bf k}$ point, the net response in the insulating phase, and the energy-resolved response.  MoS$_2$ is a nonmagnetic direct band-gap semiconductor \cite{Mak2010}, and its crystal and electronic structure are shown Fig.~\ref{fig:TMDbands}(a) and (b), respectively.  It has a direct band gap of $\Delta=1.7~{\rm eV}$ (computed value) located at ${\bf k}=\pm{\bf K}=\pm 4\pi/3a\left(1,0\right)$, where $a$ is the in-plane lattice constant. (Note that $-{\bf K}$ is often labeled ${\bf K}'$.)  Figs. 1 (c) and (d) show the ${\bf k}$-dependent Berry curvature and orbital Hall conductivity, respectively. For energies away from the band edge, the valley index of a state can be associated with the nearest valley, but the label becomes less well-defined.

The unit cell contains 2 S atoms and 1 Mo atom, and there is significant $p$-$d$ orbital hybridization in some regions of the band structure.  The states near the band gap at ${\bf K}$ and $-{\bf K}$ exhibit the largest contribution to the Berry curvature.  These states are composed primarily of $d$-orbitals localized on the Mo atoms, which form a hexagonal lattice, and which {\it do} approximately satisfy the assumptions of our analysis.  The conduction band is composed of $L_z=0$ states, corresponding to $|d_{z^2}\rangle$, while the valence band is $L_z=\pm 2$ states at ${\bf k}=\pm {\bf K}$, corresponding to $\left(|d_{x^2-y^2} \rangle \pm i |d_{xy}\rangle\right)/\sqrt{2}$ \cite{xiao2012coupled}.  These essential features are shared by other 2H TMD's \cite{xiao2012coupled}, so that our formalism is applicable to this family of materials.


In the vicinity of $\pm{\bf K}$, the conduction and valence bands are described by a massive Dirac Hamiltonian \cite{xiao2012coupled}.  Letting ${\bf q}={\bf k}\mp{\bf K}$, we have:
\begin{eqnarray}
H_\nu = t a\left( \nu q_x \tau_x + q_y \tau_y \right) + \frac{\Delta}{2} \tau_z \label{eq:HTMD}
\end{eqnarray}
where ${\bf \tau}$ is the Pauli matrix in the space of conduction and valence band states, and $\nu=\pm 1$ is the valley index.  We ignore the spin degree of freedom, as it doesn't play an essential role and makes quantitative contributions on the order of $\lambda/\Delta$, where $\lambda$ is the atomic spin-orbit parameter \cite{bhowal2020intrinsic}.  In MoS$_2$ this ratio is approximately $10^{-2}$.  For TMD's composed of heavier elements, the larger spin-orbit splitting will change some details of the analysis that we present here.

\begin{figure}
    \includegraphics[width=1.\columnwidth]{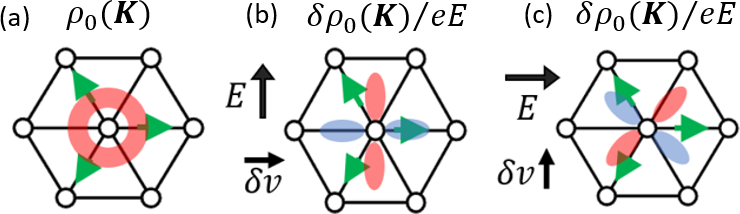}
	\caption{Depiction of how intrinsic charge current is formed at ${\bf K}$ point.  The green arrows along the 3 unique nearest-neighbor bonds show the sign of $\sin({\bf K}\cdot{\bf R})$.  In (a), the valence band unperturbed charge density is isotropic.  The bond currents add equally and sum to zero.  (b) With an E-field in the $y$-direction, the change in density (with red (blue) is positive (negative)) leads to a net current in the $x$-direction.  (c)  With an E-field in the $x$-direction, the change in density is along the 45$^\circ$ direction, leading to a net current in the $y$-direction. }\label{fig:cartoon}
\end{figure}

We first consider the currents and charge densities of the equilibrium states at ${\bf q}=0$ (or ${\bf k}={\bf K}$).  The eigenstates are $\psi_c = \left(\begin{array}{c} 1 \\ 0 \\\end{array} \right)$ and $\psi_v = \left(\begin{array}{c} 0 \\ 1 \\\end{array} \right)$.  The ${\bf R}\sin({\bf K}\cdot {\bf R})$ factor of Eq.~\ref{eq:vrho} contributes to the bond currents as indicated by the bond-aligned arrows Fig.~\ref{fig:cartoon}(a).  The charge density of both conduction and valence band equilibrium states is isotropic in the $x$-$y$ plane, so that the bond currents are weighted equally and the net current vanishes.

Applying an electric field in the $y$-direction perturbs the valence band wave function: $\psi_v' = \psi_v + (eEta/\Delta^2) \psi_c$, where $e$ is the magnitude of the electric charge.  The mixing of valence and conduction band states leads to an anisotropic $\rho({\bf r})$ which overlaps unevenly with the nearest neighbor bonds.  This in turn breaks the balance of bond currents between nearest neighbors and enables a nonzero net current, corresponding to the transverse conductivity or Berry curvature $\Omega({\bf K})$.  The modification of the density for electric fields along $x$ and $y$-directions is depicted schematically in Fig.~\ref{fig:cartoon}(b) and (c), and provides an intuitive rationale for the formation of Hall current at the ${\bf K}$ point \cite{mak2014valley,lee2016electrical}.


\begin{figure}
    \includegraphics[width=1.\columnwidth]{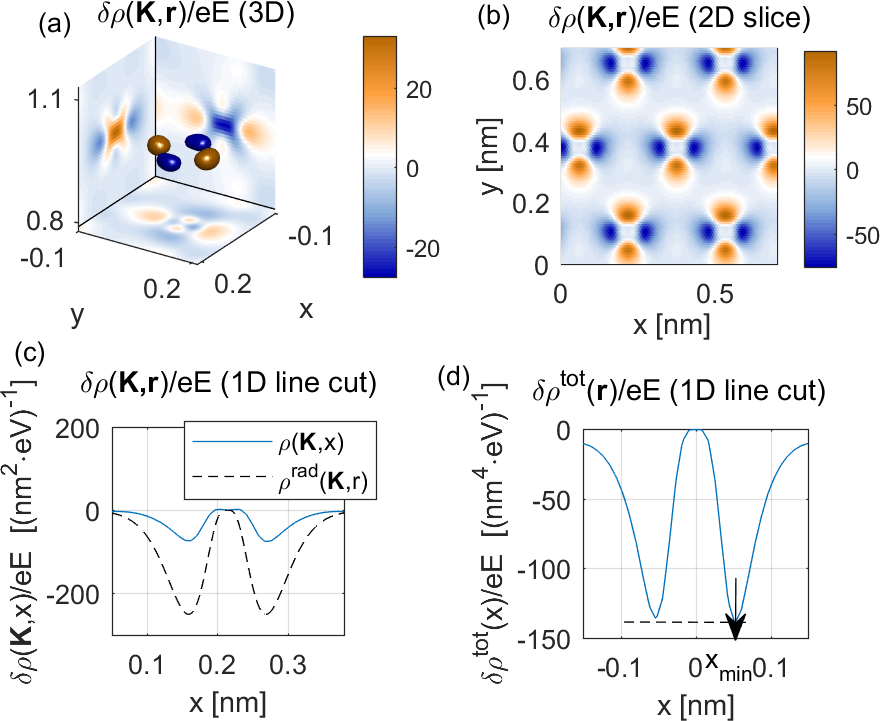}
	\caption{
(a) Isovalue contours of change in density per electric field of $\psi_v({\bf K})$ (units of $({\rm nm^2\cdot eV})^{-1}$), for an applied field in the $y$-direction.  Change of density integrated along $x$ direction is projected along the $yz$ plane (similarly for cyclic permutation of direction indices).  (b) Change of density in plane of Mo atoms.  (c) Change of density along line connecting nearest neighbors along the $x$-direction.  The normalized radial function obtained from the total change of density is also shown.  (d) Change of total density of the insulating state, along the line $y=z=0$.
}\label{fig:tmdrho}
\end{figure}

To test the quantitative validity of Eq.~\ref{eq:vrho} for MoS$_2$, we compute the electric field-induced change in charge density of the conduction band state at ${\bf K}$ using the first principles tight-binding Hamiltonian and real space orbitals obtained with Quantum Espresso \cite{QE} and Wannier90 \cite{Wannier90}.  We can independently compute both left and right-hand sides of Eq.~\ref{eq:berry3} in a model with no approximations beyond those found in density functional theory.  This provides a stringent test on the applicability of our formalism to real materials, and yields a prediction for the observable quantity $\delta\rho({\bf r})$. Fig.~\ref{fig:tmdrho}(a) shows isosurface contours of the electric field-induced change in density for the conduction band state at ${\bf k}={\bf K}$, for an applied electric field along the $y$-direction.  Fig.~\ref{fig:tmdrho}(b) shows $\delta \rho({\bf K},{\bf r})$ in the plane of the Mo atoms, and Fig.~\ref{fig:tmdrho}(c) shows the density $\delta\rho ({\bf K},x)$ along a line connecting nearest neighbors.  From the $\delta \rho({\bf K},x)$ profile, we extract the (normalized) radial part $\delta \rho^{\rm rad}({\bf K},r)$ to obtain $\delta \rho^{\rm ang}({\bf K},{\bf \hat {R}})$.  Applying Eq.~\ref{eq:vrho} leads to
\begin{eqnarray}
\Omega({\bf K}) = -a \tilde{t}^\sigma \frac{\delta\rho^{\rm ang}({\bf K},{\bf {\hat x}})}{eE}
\end{eqnarray}
We use a value of $t^{\sigma}=1.09~{\rm eV}$ obtained by fitting the conduction/valence bands to the effective model near ${\bf K}$.  Plugging in numbers we obtain a predicted value $\Omega({\bf K})=0.074~{\rm nm^2}$ from the real space density analysis, compared to the directly computed value $\Omega({\bf K})=0.094~{\rm nm}^2$.  We find semi-quantitative agreement between the two values, validating the applicability of our approach to real materials.  The difference between the two quantities is traced back to the nonzero contribution of the $p$ orbitals of the wave function at ${\bf K}$.

We next extend our analysis to the full, ${\bf k}$-integrated response for the insulating system.  The Berry curvature at finite $q$ is given by:
\begin{eqnarray}
\Omega_\nu(q)&=&\nu \frac{ 2 \Delta t^2 a^2}{\left(4q^2t^2 + \Delta^2\right)^{3/2}}~.
\end{eqnarray}
The valley Hall conductivity $\sigma_{\rm VHE}$ of the effective model is obtained by integrating over $q$ and summing over $\nu$, which results in $\sigma_{\rm VHE} = \sigma_{\rm OHE} = 1/(2\pi)$.  (We omit factors of $e^2/\hbar$ in reporting conductivity values.)  The values for $\sigma_{\rm VHE}$ and $\sigma_{\rm OHE}$ obtained with first principles calculations, which are given by $0.71/(2\pi)$ and $1.05/(2\pi)$, respectively. The deviations between conductivities obtain with the effective model and first principles reflect the difference between the two models' band structure. In the Supp. Info, we show that the $q$-dependent perturbed density is proportional to the Berry curvature:
\begin{eqnarray}
\frac{\delta \rho_\nu(q,{\bf r})}{eE} &=&  \nu \frac{\Omega_\nu(q)}{\sqrt{2} t a}~\phi_{z^2}({\bf r})\phi_{x^2-y^2}({\bf r}).\label{eq:drhoq}
\end{eqnarray}
where $\phi_{z^2(x^2-y^2)}$ is the $d$-like atomic wave function for the conduction (valence) band state. Importantly, the change in charge density and the Berry curvature have the same $q$-dependence.  Integrating Eq.~\ref{eq:drhoq} over $q$ relates the total change in charge density $\delta\rho^{\rm tot}$ to the valley Hall conductivity.  Fig.~\ref{fig:tmdrho}(d) shows the line cut of the total change in charge density along $x$.  We can derive the following relation between the extrema of the change in charge density along $x$ and valley Hall conductivity:
\begin{eqnarray}
-f ta x_{\rm min}^3 \frac{\delta \rho^{\rm tot}(x_{\rm min})}{eE}&=&\sigma_{\rm VHE} \label{eq:rhoVHE}
\end{eqnarray}
where $f=3 \pi/32\sqrt{3/2} \exp(4)$.  Using the change in density obtained from first principles and Eq. \ref{eq:rhoVHE}, we obtain an estimated value for $\sigma_{\rm VHE}$ of $0.95/(2\pi)$.  This compares well with the directly computed value of valley Hall conductivity. The comparison with the orbital Hall conductivity is more favorable, which is a consequence of the close relationship between the valley and orbital Hall conductivity in this material, and because the orbital Hall conductivity is more strongly concentrated near the $\pm{\bf K}$ points (see Fig.~\ref{fig:TMDbands}(d)). The good agreement between the directly computed $\sigma_{\rm VHE}$ and the value predicted with our real space analysis again demonstrates the applicability of our approach to real materials. \\

\begin{figure}
	\includegraphics[width=1.\columnwidth]{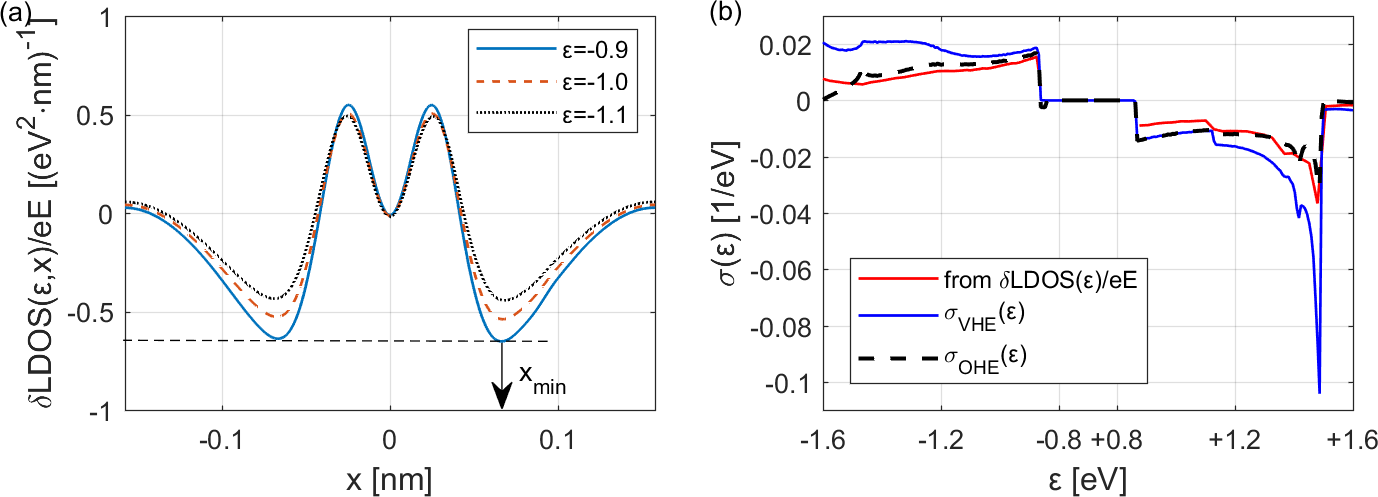}
	\caption{(a) The change in local density of states $\delta{\rm LDOS}(\varepsilon,{\bf r})/eE$ integrated along the $z$-direction for three energy values. The important features are the position of $x_{\rm min}$ and $\delta {\rm LDOS}(\varepsilon,x)/eE$ at this position. (b) Energy-resolved valley Hall (black dashed line) and orbital Hall (blue line) conductivity near the valence and conduction band edges.  The red line shows the value inferred from the local density of states measurement using Eq.~\ref{eq:omegaE} (note break in the $x$-axis).
	}\label{fig:omegaE}
\end{figure}

{\it Valley/orbital Hall spectroscopy} -- We next consider the energy-resolved valley/orbital Hall conductivities and change in density.  We again focus on a linecut along the $x$-direction, and assume that the measured local density of states is proportional the integral of the density along the $z$ direction ({\it e.g.}, the integral of Eq.~\ref{eq:drhoq} along $z$).  In the Supp. info we show that:
\begin{eqnarray}
-\int dz~\frac{\delta \rho_\nu(q,x_{\rm min},0,z)}{eE} \times \frac{ta x_{\rm min}^2}{0.065} &=&  \nu ~\Omega_\nu(q)
\end{eqnarray}
where $x_{\rm min}$ is the position of the minimum in the change of density (see Fig.~\ref{fig:omegaE}(a)).  The proportionality between the $q$-dependent change in charge density and the Berry curvature implies that the two quantities are also proportional when integrating over $q$.  In particular, the energy-resolved quantities are proportional.  Denoting the energy-resolved density, or local density of states, as ${\rm LDOS}$, we find:
\begin{eqnarray}
-\frac{ta x_{\rm min}^2}{0.065} \left( \frac{\delta {\rm LDOS}(\varepsilon,x_{\rm min})}{eE} \right) &=& \sigma_{\rm VHE}(\varepsilon)\label{eq:omegaE}
\end{eqnarray}

Fig.~\ref{fig:omegaE}(a) shows the energy-resolved change in density for three energies below the valence band edge. Fig.~\ref{fig:omegaE}(b) shows a comparison between the directly computed valley Hall/orbital conductivity and the predicted value based on Eq.~\ref{eq:omegaE}.  We see semi-quantitative agreement up to 500~meV away from conduction and valence band edges.  We observe that the change in density adheres more closely to the orbital Hall effect at energies away from the band edges.  This is due to the fact that the change in density and orbital Hall effect are dominated by states near $\pm {\bf K}$ for most energies, whereas the valley Hall conductivity acquires contributions from states near $\Gamma$ (${\bf k}=0$) at energies away from the band edges.

To summarize, we show that the real space density response of TMD's such as MoS$_2$ provides a direct estimate of the intrinsic valley Hall conductivity.  This is a consequence of a relation we derive between the Berry curvature and the charge density response of a specific class of lattices.  Experimentally, various probes attain the required sub-Angstrom spatial resolution, such as X-ray diffraction \cite{wahlberg2015powder}, electron beam diffraction \cite{gao2019real}, and scanning tunneling microscopy \cite{jelinek2017high}.  We estimate a fractional change of the local density of states on the order of $10^{-4}$ at the breakdown electric field of MoS$_2$ \cite{lembke2012breakdown}. The measurement can be calibrated with the ground state density or LDOS, as described in the Supp. Info, enabling the quantitative estimate of the valley Hall conductivity.

F.X. acknowledges support under the Cooperative Research Agreement between the University of Maryland and the National Institute of Standards and Technology Physical Measurement Laboratory, Award 70NANB14H209, through the University of Maryland.

\bibliography{TMD}
\bibliographystyle{apsrev4-1}

\pagebreak[4]

\onecolumngrid
\vskip 0.55in
\begin{center}
  \Large\textbf{Supplementary Information for ``Imaging the valley and orbital Hall effect in monolayer MoS$_2$''}
\end{center}

\section{Relation Between density and current}

In this section we derive the derivation of the relation between a state's charge density distribution and its velocity in a tight-binding model.  This relation holds only for a lattice composed of a monoatomic unit cell, with orbitals of the same character ({\it e.g.}, $p$ or $d$ orbitals), and a crystal field Hamiltonian with only $\sigma$-hopping.  This analysis is therefore qualitatively valid for cases in which $t^\pi,t^\delta$ are sufficiently less than $t^\sigma$ bonding.  This is often the case, and in the next section we provide an explicit description of the validity of this assumption.\\

The tight-binding model is represented in real-valued spherical harmonics.  We denote these by $\rYlm$, which are linear combinations of $Y_\ell^m$ and $Y_\ell^{-m}$.  Letting  $m=|M|$:
\begin{eqnarray}
\rYlm = \begin{cases} \frac{1}{\sqrt{2}}\left(Y_\ell^{-m} + (-1)^m Y_\ell^{m}\right)~~~~{M<0},\\
        Y_\ell^0~~~~~~~~~~~~~~~~~~~~~~~~~~~~~~~~{M=0},\\
        \frac{i}{\sqrt{2}}\left(Y_\ell^{-m} - (-1)^m Y_\ell^{m}\right)~~~~{M>0}~.
        \end{cases}
\end{eqnarray}
The specific forms of the real spherical harmonics for a given $\ell$ can be found in various references.  Generally the $M$ label is expressed in terms of the cartesian factors corresponding to the form of the spherical harmonic ({\it e.g.} for $\ell=1,~M=\{-1,0,1\}$ is labeled as $\{x, z, y\}$).

\begin{figure}
    \includegraphics[width=.75\columnwidth]{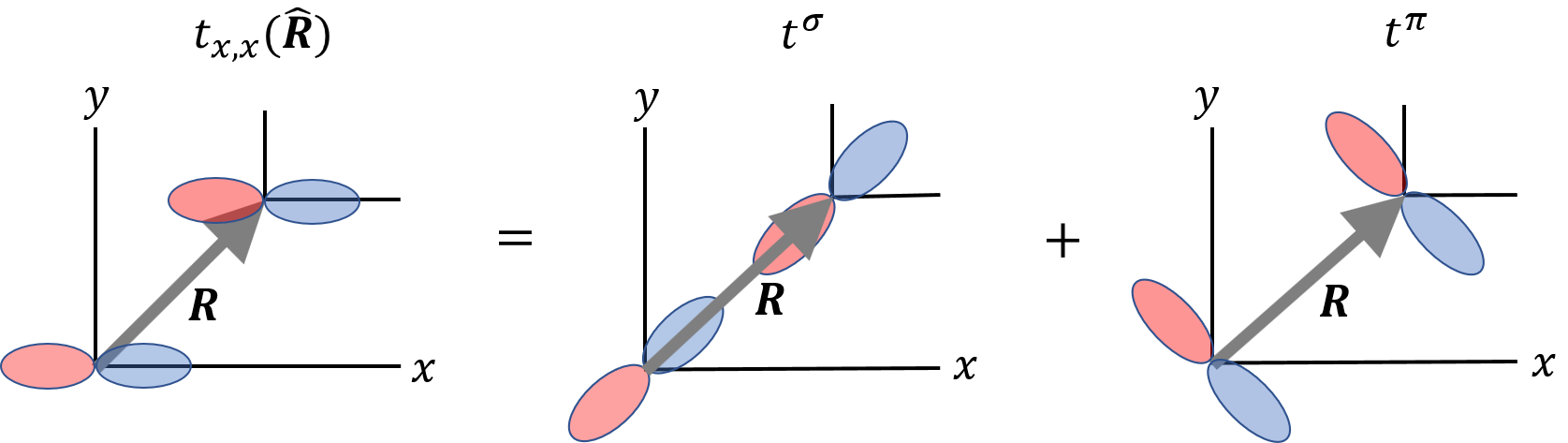}
	\caption{Depiction of the how hopping between a pair of $p_x$ orbitals is decomposed into $\sigma$ and $\pi$ contributions.}\label{fig:SK_TB}
\end{figure}

In the tight-binding formulation, the $\sigma$-bonding hopping parameter $t^\sigma$ is defined as the hopping between $m=0$ orbitals displaced along the $z$-direction.
\begin{eqnarray}
t^\sigma(R) = \int d{\bf r}~V({\bf r})~ Y_\ell^0({\bf r})~ Y_\ell^0\left({\bf r}-R {{\bf \hat z}}\right) R_{n,\ell}({\bf r})~R_{n,\ell}({\bf r}-R{\bf z})\label{eq:tsigma}
\end{eqnarray}
The $t^\sigma$ hopping between orbital $\alpha$ and $\beta$ displaced along the $\hat n$ direction is then determined by projecting each orbital along the $|\ell,m=0\rangle_{\hat n}$ (see Fig. \ref{fig:SK_TB}(a) and (b)).  This projection is obtained by rotating the $|\ell, m=0\rangle_{\hat z}$ into the $\hat n$-direction (see Fig. \ref{fig:SK_TB}) We review this procedure next.\\

We denote the axis of quantization $\hat n$ for the spherical harmonic in the subscript of the ket: $|\ell, m\rangle_{\hat n}$.  The operator which rotates $|\ell, m\rangle_{\hat z}$ into $|\ell, m \rangle_{\hat n}$ is denoted by $R(\hat n)$:
\begin{eqnarray}
|\ell, m \rangle_{\hat n} = R(\hat n) |\ell, m \rangle_{\hat z}
\end{eqnarray}
The rotated spherical harmonic $|\ell, m \rangle_{\hat n}$ can be written as a linear combination of unrotated spherical harmonics with the same value of $\ell$, written in terms of the Wigner $D$-matrix:
\begin{eqnarray}
R(\hat n) |\ell, m\rangle_{\hat z} = |\ell, m\rangle_{\hat n} = \sum_{m'} D^\ell_{m,m'}(\hat n) |\ell, m'\rangle_{\hat z}
\end{eqnarray}
Crucially, for $m=0$ the elements of $D_{m=0,m'}^\ell$ are equal to the value of the spherical harmonic $|\ell, m'\rangle$ evaluated at $\hat n=\left(\theta,\phi\right)$.
\begin{eqnarray}
|\ell, m=0\rangle_{\hat n} = \sum_m \left(Y^m_\ell\left(\theta,\phi\right)\right)^* |\ell, m \rangle_{\hat z} \label{eq:rotatem0}
\end{eqnarray}
The projection of $|\ell,m=0\rangle_{\hat n}$ onto orbital $|\ell, m\rangle_{\hat z}$ is then given by Eq. \ref{eq:rotatem0}.  As described in the previous paragraph, this quantity provides the angular dependence of the hopping between orbitals displaced by ${\bf R}$.
\begin{eqnarray}
_{\hat z}\langle \ell,m | \ell, m=0 \rangle_{\hat n} = \left(Y_\ell^m(\theta,\phi)\right)^*
\end{eqnarray}

The $\sigma$-hopping contribution between orbital $\alpha$ and $\beta$ displaced by ${\bf R}$ is therefore equal to $y_\alpha({\bf \hat{R}})$ and $y_\alpha({\bf \hat{R}})$:
\begin{eqnarray}
t_{M,M'}^\sigma({\bf {\hat R}}) = t^\sigma\left(\frac{\rYlm({\bf {\hat R}})~ {\mathcal Y}_{\ell}^{M'}({\bf {\hat R}})}{{\mathcal Y}_{\ell}^{0}({\bf {\hat z}})^2}\right) .\label{eq:trho}
\end{eqnarray}
The denominator of Eq. \ref{eq:trho} is a normalization factor that ensures that Eq. \ref{eq:tsigma} is recovered for hopping between $m=0$ orbitals displaced along the $z$-direction ({\it i.e.,} for $\theta=\phi=0$).

This form of $t$ leads to the relation between current and density.  We write the general form of the current:
\begin{eqnarray}
{\bf v}({\bf k})=-\sum_{{\bf R},M,M'}{\bf R}\sin({\bf k}\cdot{\bf R})~t_{M,M'}^\sigma({\bf R})\left(c_M^{\dagger}c_{M'}+h.c.\right)\label{eq:v}
\end{eqnarray}
Next we write the equation for the density from primary unit cell orbitals:
\begin{eqnarray}
\rho_0({\bf r}) &=& \sum_{M,M'} \left(c_M^{\dagger}c_{M'}+h.c.\right) ~\rYlm\left({\bf {\hat r}}\right){\mathcal Y}_{\ell}^{M'}\left({\bf {\hat r}}\right) R_{n,\ell}(r) R_{n,\ell}(r)  \\
 &=&R_{n,\ell}(r)^2 \sum_{M,M'} \left(c_M^{\dagger}c_{M'}+h.c.\right) ~\rYlm\left({\bf {\hat r}}\right){\mathcal Y}_{\ell}^{M'}\left({\bf {\hat r}}\right)
\end{eqnarray}
As described in the main text, the charge density is the product of a radial function and an angular function
\begin{eqnarray}
\rho_0\left({\bf r}\right) = \rho_0^{\rm rad}(r) \rho_0^{\rm ang}\left({\bf {\hat r}}\right)
\end{eqnarray}
where $\rho_0^{\rm rad}(r)=R_{n,\ell}(r)^2$ and
\begin{eqnarray}
\rho_0^{\rm ang}\left({\bf {\hat r}}\right) = \sum_{M,M'} \left(c_{M}^{\dagger}c_{M'}+h.c.\right) ~\rYlm\left({\bf {\hat r}}\right){\mathcal Y}_{\ell}^{M'}\left({\bf {\hat r}}\right). \label{eq:rho}
\end{eqnarray}
Combining Eqs. \ref{eq:trho},~\ref{eq:v},~and \ref{eq:rho}, we obtain:
\begin{eqnarray}
{\bf v}({\bf k})\propto -\sum_{{\bf R}} \frac{t^\sigma(R)}{Y_\ell^0(0,0)^2}~ {\bf R}\sin({\bf k}\cdot{\bf R})~\rho^{\rm ang}_0({\bf {\hat R}})
\end{eqnarray}
$t^\sigma(R)$ is the $R$-dependent value of the sigma-bonding hopping integral, whose generally parameterized form can be found in the following section.

\section{Limit of validity for $\sigma$-hopping tight-binding}

\begin{figure}
    \includegraphics[width=1.\columnwidth]{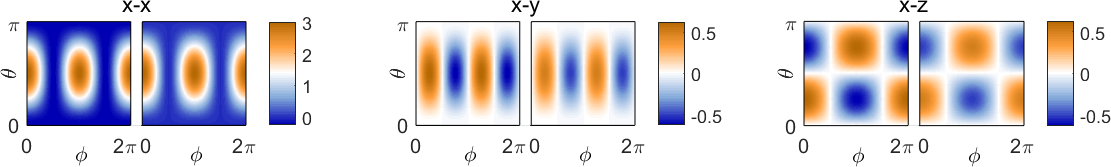}
	\caption{Comparison of the total hopping (left side of each subplot) to $\sigma$-hopping (right side of each subplot) as a function of the relative orientation (parameterized by polar angles $(\theta,\phi)$ for all unique pairs of $p$-orbitals.  }\label{fig:hoppingp}
\end{figure}

\begin{figure}
    \includegraphics[width=1.\columnwidth]{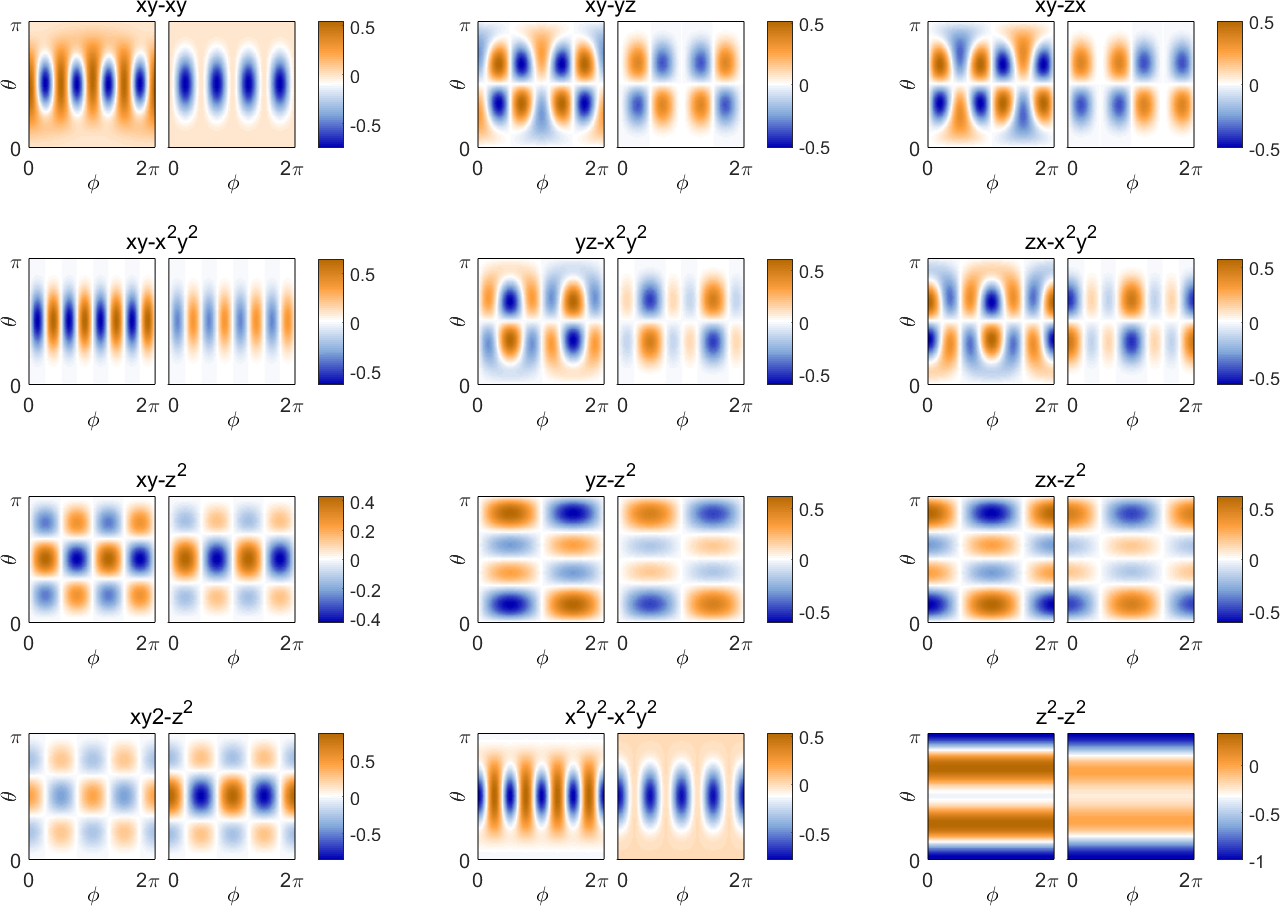}
	\caption{Comparison of the total hopping (left side of each subplot) to $\sigma$-hopping (right side of each subplot) as a function of the relative orientation (parameterized by polar angles $(\theta,\phi)$) for all unique pairs of $d$-orbitals. }\label{fig:hoppingd}
\end{figure}

The previous derivation applies for $\sigma$-hopping.  Generally $\pi$ and $\delta$-hopping can be quantitatively, and even qualitatively as important as $\sigma$-hopping.  The relative importance of different hopping terms depends on the type of inter-orbital hopping and the relative orientation of the two orbitals.  Fig. \ref{fig:hoppingd} shows both the $\sigma$-hopping and total hopping amplitudes for all of the unique interorbital hopping for $\ell=1$, as a function of relative orientation (parameterized by polar angles $(\theta,\phi)$).  The form of the tight-binding matrix elements are taken from \cite{harrison2012electronic}.  For $p$-orbitals, the parameterization is:
$V_{pp\sigma}=3.24,~V_{pp\pi}=-0.81$, and the hopping is \cite{harrison2012electronic}:
\begin{eqnarray}
t_{pp\alpha}(R) = V_{pp\alpha} \frac{\hbar^2}{2 m R^2} \label{eq:tpp}
\end{eqnarray}
Fig. \ref{fig:hoppingp} shows that for $p$-orbitals, the $\sigma$-hopping is always a good semi-quantitative estimate of the total hopping for all configurations.

For $\ell=2$ ($d$-orbitals), the parameters are taken to be $V_{dd\sigma}=-16.2,~V_{dd\pi}=8.75,~V_{dd\delta}=-2.3$, and the hopping is \cite{harrison2012electronic}:
\begin{eqnarray}
t_{dd\alpha}(R) = V_{dd\alpha} \frac{\hbar^2 R_d^3}{2 m R^5}
\end{eqnarray}
where $R_d$ is another parameter, and is typically element-specific.  A similar conclusion of the predominance of $\sigma$-hopping is reached for $d$ orbitals(Fig.~\ref{fig:hoppingd}), with one notable exception: the $d_{zx}-d_{x^2y^2}$ total hopping and $\sigma$-hopping are off by a minus sign for most orbital configurations.  However for all other orbital pairs, the $\sigma$-hopping at least qualitatively, and often semi-quantitatively provides a representation of the total hopping.

\section{Real space density calculations}

We describe the procedure used to obtain the real space density associated with the perturbed eigenstates.  We first obtain the real space orbitals from Wannier90, extending out 2 unit cells away from the atom center $\phi_\alpha({\bf r})$ (so that the density is represented in a $5 \times 5$ supercell).  The wave function is expressed as a linear combination of these basis orbitals:
\begin{eqnarray}
| \psi_i \rangle &=& \sum_\alpha c_\alpha({\bf k})  | \phi_\alpha ({\bf r}) \rangle
\end{eqnarray}
The ${\bf k}$-dependence of the wave function coefficients is given by Bloch phase factors. We form the perturbed state as:
\begin{eqnarray}
|\psi_i\rangle' = |\psi_i \rangle + i eE \sum_{j\neq i} \frac{\langle \psi_j | \frac{dH}{dk_x} | \psi_i \rangle}{\left(E_i - E_j\right)^2} | \psi_j \rangle
\end{eqnarray}
where $E$ is a small parameter.

Given a set of occupied states, the associated density matrix is given by the outer product of the states:
\begin{eqnarray}
\rho &=& \sum_{i} f_i | \psi_i \rangle \langle \psi_i |
\end{eqnarray}
The real space density associated with the density matrix is:
\begin{eqnarray}
\rho({\bf k} ,{\bf r}) = \sum_{\alpha,\beta,{\bf R},{\bf R'}}  \rho_{\alpha,\beta}~ \phi_\alpha\left({\bf r} - {\bf R}\right) \phi_\beta \left({\bf r} - {\bf R'}\right) \exp\left(i{\bf k}\cdot \left({\bf R} - {\bf R'}\right)\right) \label{eq:denstotal}
\end{eqnarray}

\section{Relation between total density versus primary unit cell density}

\begin{figure}
    \includegraphics[width=0.7\columnwidth]{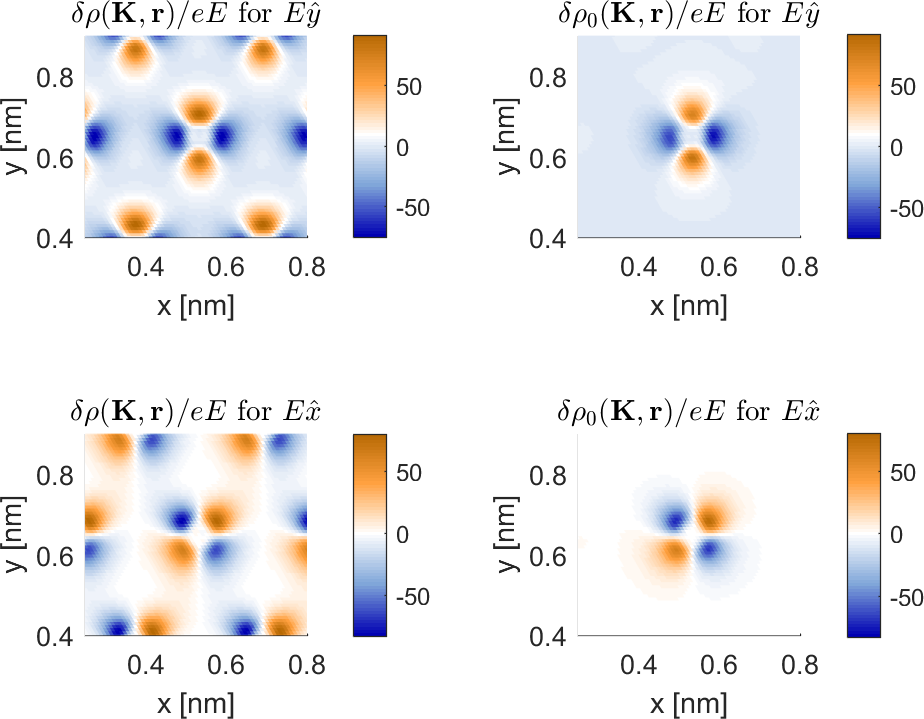}
	\caption{Comparison between electric field-induced change of charge density for all charge ($\delta\rho$ of Eq.~\ref{eq:denstotal}), versus charge only from ``on-site'' contributions ($\delta\rho_0$ of Eq.~\ref{eq:rho}) for electric fields aligned to the ${\bf x}$ and ${\bf y}$ directions in monolayer MoS2.  The difference between the two is minimal near atomic centers. Units for all figures are $({\rm eV\cdot nm^2})^{-1}$.}\label{fig:deltarho0}
\end{figure}

The density which enters into the relation between velocity and charge density is not the total density, but the density contribution from ``on-site'' orbitals $\rho_0({\bf r})$.  The operator form is as given in the main text:
\begin{eqnarray}
\rho_0({\bf r}) &=& \sum_{M,M'} \left(c_{M}^{\dagger}c_{M'}+h.c.\right) ~\phi_M\left({\bf r}\right)
    \phi_{M'}\left({\bf r}\right) \label{eq:rho0}
\end{eqnarray}
In terms of the density matrix, this quantity is:
\begin{eqnarray}
\rho_0({\bf r}) = \sum_{\alpha,\beta}  \rho_{\alpha,\beta}~ \phi_\alpha\left({\bf r}\right) \phi_\beta \left({\bf r}\right)
\end{eqnarray}
$\rho_0$ corresponds to only including terms ${\bf R}={\bf R'}=0$ in Eq. \ref{eq:denstotal}.  The charge density in the interstitial regions is quite different for $\rho({\bf r})$ and $\rho_0({\bf r})$.  However in the vicinity of the nuclei, these densities are quite similar.  We show this explicitly in Fig.~\ref{fig:deltarho0}, where we plot $\delta\rho({\bf r})/eE$ and $\delta\rho_0({\bf r})/eE$ for applied field in ${\bf x}$ and ${\bf y}$ directions in MoS2.  As expected, the quantitative difference is quite small near the nuclei.  This can be generally anticipated for states with high Berry curvature:  Berry curvature is related to the formation of orbital moments, which requires higher values of $\ell$.  These states are more localized than, for example, $s$-like states.


\section{Real space calculation for MoS$_2$}

We begin our analysis of MoS$_2$ by considering the effective model near the $\pm{\bf K}$ points.  We let ${\bf q}=({\bf k}\mp {\bf K})$ be the Bloch momentum vector measured from the $\pm {\bf K}$ point.  The effective model Hamiltonian is:
\begin{eqnarray}
H_{\nu=\pm}&=&\left(\begin{array}{cc}
\Delta / 2 & \pm tqa e^{\mp i \phi} \\
\pm tqa e^{\pm i \phi} & -\Delta / 2
\end{array}\right)
\end{eqnarray}
where $\nu=\pm 1$ is the valley index and $a$ is the lattice constant.  The eigenvalues for both valleys are given by:
\begin{eqnarray}
\varepsilon(q)&=&\pm 1 / 2\left(\Delta^{2}+4 q^{2}a^2 t^{2}\right)^{1 / 2} .
\end{eqnarray}
It's straightforward to show that the Berry curvature is given by:
\begin{eqnarray}
\Omega_\nu(q) &=&\nu\frac{2 t^{2} a^2 \Delta}{\left(4 t^{2} q^{2}a^2+\Delta^{2}\right)^{3 / 2}}~. \label{eq:Omegaq}
\end{eqnarray}
\\

We next record the full form of the wave functions of conduction and valence bands (denoted by $\psi_c$ and $\psi_v$, respectively) at ${\bf k}=+{\bf K}$, which form the Hilbert space for nonzero $q$:
\begin{eqnarray}
\phi_c(x,y,z) &=& \phi_{z^2} \\
&=&A_r \exp\left(\frac{-r}{3a_0}\right)  A_{1} \left(-x^2 - y^2 + 2z^2\right) \\
\phi_v(x,y,z) &=& \frac{1}{\sqrt{2}}\left(\phi_{x^2-y^2} + i \phi_{xy}\right) \\
&=& A_r \exp\left(\frac{-r}{3a_0}\right) \frac{1}{\sqrt{2}}\left(i A_{2} \left(xy\right) +  A_{3} \left(x^2 - y^2\right)\right)
\end{eqnarray}
where $r=\sqrt{x^2 + y^2 + z^2}$.  $a_0$ is the effective Bohr radius of the orbital.  The various normalization factors $A$ are:
\begin{eqnarray}
A_r &=& \left(\frac{2}{27}\sqrt{\frac{2}{5}} \left(\frac{1}{3a_0}\right)^{3/2} \frac{1}{a_0^2}\right)\\
A_{1} &=& \frac{1}{4}\sqrt{\frac{5}{\pi}},~~A_{2} = \frac{1}{2}\sqrt{\frac{15}{\pi}}, ~~A_{3} = \frac{1}{4}\sqrt{\frac{15}{\pi}}
\end{eqnarray}
Notice that the radial wave function has a prefactor of $r^2$ in the numerator, while the spherical harmonic has a factor of $r^2$ in the denominator, the two of which cancel each other.\\

To determine the form of the perturbed wave functions, and resulting real space density away from the $+{\bf K}$ point ($\nu=+1$), we return to the effective model:

\begin{eqnarray}
H&=&\left(\begin{array}{cc}
\Delta / 2 & tqa e^{-i \phi} \\
tqa e^{i \phi} & -\Delta / 2
\end{array}\right)
\end{eqnarray}

The unperturbed eigenstates are spinors oriented along the direction of the pseudo effective field ${\bf B}_{\rm eff}~=~\left(2tqa \cos(\phi),2tqa \sin(\phi),\Delta\right)$.
\begin{eqnarray}
\psi_c(q) &=& \left(\begin{array}{c}
\cos(\theta/2) e^{-i\phi}\\
\sin(\theta/2) 
\end{array}\right)\\
\psi_v(q) &=& \left(\begin{array}{c}
\sin(\theta/2) e^{-i\phi}\\
-\cos(\theta/2) 
\end{array}\right)
\end{eqnarray}
where $\theta = \tan^{-1}\left(\frac{2qat}{\Delta}\right)$.  Recall that the psuedospin basis functions are combinations of the $d$-orbitals:
\begin{eqnarray}
\left(\begin{array}{c}
1\\
0
\end{array}\right) &=& d_{z^2} \\ &=&A_1\left(2z^2 - x^2 - y^2\right) \\
\left(\begin{array}{c}
0\\
1
\end{array}\right) &=& \frac{1}{\sqrt{2}}\left(d_{x^2-y^2} + i d_{xy} \right)\\ &=& \left( A_3 (x^2-y^2)  + i A_2 xy\right)
\end{eqnarray}
The perturbed wave function is:
\begin{eqnarray}
|\psi_v'\rangle = |\psi_v\rangle + ieE\frac{\langle \psi_c |v_y | \psi_v\rangle}{(E_c-E_v)^2} |\psi_c \rangle
\end{eqnarray}
Let's consider $\phi=0$.
\begin{eqnarray}
|\psi_v'\rangle = \left(\begin{array}{c}
\sin(\theta/2) \\
-\cos(\theta/2)
\end{array}\right) + \frac{eaEt}{4q^2a^2t^2+\Delta^2}\left(\begin{array}{c}
\cos(\theta/2) \\
\sin(\theta/2)
\end{array}\right)
\end{eqnarray}

If we evaluate the 1$^{\rm st}$ order change in the wave function, we find that following expression, where we include the functional form of the angular part of the orbitals:
\begin{eqnarray}
\frac{\delta \rho_{\nu=+1}(q,{\bf r})}{eE} &=& \frac{ta\left(2 qa t (|\phi_v({\bf r})|^2 - |\phi_c({\bf r})|^2)+2\Delta~ \phi_c({\bf r})~{\rm Re}\left[\phi_v({\bf r})\right]\right)}{\left(4q^2a^2t^2+\Delta^2\right)^{3/2}}
\end{eqnarray}
Since $qta\ll \Delta$, we can approximate the above as:
\begin{eqnarray}
\frac{\delta \rho_{\nu=+1}(q,{\bf r})}{eE}&\approx& \frac{2ta\Delta~ \phi_c({\bf r})~ {\rm Re}\left[\phi_v({\bf r})\right]}{\left(4q^2a^2t^2+\Delta^2\right)^{3/2}}\\
&=& \frac{2ta\Delta}{\left(4q^2a^2t^2+\Delta^2\right)^{3/2}}~ \frac{\phi_{z^2} ({\bf r}) \phi_{x^2-y^2}({\bf r})}{\sqrt{2}} \label{eq:deltarho}
\end{eqnarray}
Note that the unit for $\delta \rho(q,{\bf r})$ is the standard unit for density, ${\rm 1/m^3}$.  We find an identical expression for $\nu=-1$, so that the net density is nonzero when summing over valleys.


\section{Insulating case: relating $\sigma_{\rm VHE}$ and $\delta\rho^{\rm tot}({\bf r})/eE$}

We integrate over ${\bf k}$ (or ${\bf q}$) to find the total valley Hall conductivity in the insulating case:
\begin{eqnarray}
\sigma_{\rm VHE} &=& \sum_{\nu=\pm 1}\frac{ 2\pi}{(2\pi a)^2} \int_0^\infty dq~q~  \nu~ \Omega_{\nu} (q)\label{eq:VHE_insulator0} \\
&=& \frac{4\pi}{(2\pi a)^2} \int_0^\infty dq~q~ \frac{2 t^{2}a^2 \Delta}{\left(4 t^{2} q^{2}a^2+\Delta^{2}\right)^{3 / 2}} \label{eq:VHE_insulator}\\
&=& \frac{1}{2\pi}
\end{eqnarray}
The total (integrated) change in density is given by:
\begin{eqnarray}
\frac{\delta \rho^{\rm tot}({\bf r})}{eE} = \frac{\phi_{z^2}({\bf r}) \phi_{x^2-y^2}({\bf r})}{\sqrt{2}}  \frac{2\times 2\pi}{(2\pi a)^2}\int q~dq~\frac{2 a t\Delta }{\left(4q^2a^2t^2 + \Delta^2\right)^{3/2}} ~.
\end{eqnarray}
A factor of 2 in the above is the result of summing over $\nu$.  Note the units of $\rho({\bf r})^{\rm tot}$, the ``total'' density, are ${\rm 1/m^5}$, which include a factor of ${\rm 1/m^3}$ from the standard density unit, and an additional factor ${\rm 1/m^2}$ from the Brillouin zone integration.

Next we connect the two quantities by picking out a specific point ${\bf r}$.  The natural point is the maximum of the density along the $x$-direction, for $y=z=0$.  Letting $y=z=0$:
\begin{eqnarray}
\frac{\delta \rho^{\rm tot}(x)}{eE} = -\frac{\exp\left(-\frac{2x}{3a_0}\right)x^4}{13122\sqrt{6}a_0^7 \pi}\frac{4\pi}{(2\pi a)^2}\int q~dq~\frac{2 a t\Delta }{\left(4q^2a^2t^2 + \Delta^2\right)^{3/2}} ~.
\end{eqnarray}
In the above $a_0$ is the effective Bohr radius  The minimum value of $\rho(x)$ is located at $x=6a_0$.  At this point:
\begin{eqnarray}
\frac{\delta \rho^{\rm tot}(x_{\rm min})}{eE} =
-\frac{32}{3\pi}\sqrt{\frac{2}{3}} ~\frac{\exp(-4)}{x_{\rm min}^3} \left[\frac{4\pi}{(2\pi a)^2}\int q~dq~\frac{2  at \Delta}{\left(4q^2a^2t^2 + \Delta^2\right)^{3/2}} \right]~.
\end{eqnarray}
We use Eq. \ref{eq:VHE_insulator} to rewrite the term in brackets on the right-hand-side of the above as:
\begin{eqnarray}
\frac{\delta \rho^{\rm tot}(x_{\rm min})}{eE} =
-\frac{32}{3\pi}\sqrt{\frac{2}{3}}~ \frac{\exp(-4)}{x_{\rm min}^3}\left( \frac{\sigma_{\rm VHE}}{at} \right)~,
\end{eqnarray}
equivalently:
\begin{eqnarray}
\sigma_{\rm VHE} =
-\frac{3\pi}{32}\sqrt{\frac{3}{2}}~ \exp(4)~x_{\rm min}^3 at \frac{\delta \rho(x_{\rm min})}{eE}~.\label{eq:VHE_RDS}
\end{eqnarray}
We can test this picture by separately computing the two sides of Eq. \ref{eq:VHE_RDS}, which is presented in the main text.

\section{Change in local density of states}

For the local density of states calculation, we focus on $y=0$ and will integrate the density over $z$. Re-writing Eq. \ref{eq:deltarho}:
\begin{eqnarray}
\frac{\delta\rho(q,x,0,z)}{eE} = \frac{ 2 \Delta at }{\left(4q^2a^2t^2 + \Delta^2\right)^{3/2}}\frac{A_r^2 A_1 A_3}{\sqrt{2}} \exp\left(\frac{-2\sqrt{x^2+z^2}}{3a_0}\right)x^2 \left(-x^2+2z^2\right)
\end{eqnarray}
Here we omit the $\nu$ label for $\delta\rho(q,{\bf r})$, as the density is equal for both $\nu$ values as described earlier.  Generally, a two-dimensional map of charge density obtained with scanning tunneling microscopy (STM) involves a convolution over the depth $z$.  The precise form of this convolution function $f(z)$ depends on the experimental details.  For the sake of simplicity, we choose $f(z)=1$ in the analysis presented here.  Generalizing to other forms of $f(z)$ is straightforward, and will result in different numerical prefactors whose precise value is important for quantitative data analysis.  Proceeding with $f(z)=1$, we obtain:
\begin{eqnarray}
\int dz~\frac{\delta \rho(q,x,0,z)}{eE} &=&  \frac{ 2 \Delta a t }{\left(4q^2a^2t^2 + \Delta^2\right)^{3/2}} \int dz\frac{1}{\sqrt{2}}  A_r^2 A_1 A_3 \exp\left(\frac{-2\sqrt{x^2+z^2}}{3a_0}\right)x^2 \left(-x^2+2z^2\right)  \\
\end{eqnarray}
To determine the position of the absolute value of the maximum of this function, we first make integral dimensionless with $z' = z/x$.  We also include the numerical prefactors explicitly:
\begin{eqnarray}
\int dz~\frac{\delta \rho(q,x,0)}{eE} &=&   \frac{ 2 \Delta a t }{\left(4q^2a^2t^2 + \Delta^2\right)^{3/2}}\left(\frac{1}{27^3}\right) \sqrt{\frac{3}{2}}\frac{1}{2\pi} \frac{1}{a_0^7} \int x~ dz'  \exp\left(\frac{-2x\sqrt{1+z'^2}}{3a_0}\right)x^4 \left(-1+2z'^2\right)
\end{eqnarray}
Next write $x$ in dimensionless form, $x'=x/a_0$.
\begin{eqnarray}
\int dz~\frac{\delta \rho(q,x',0,z)}{eE} &=&  \frac{ 2 \Delta a t }{\left(4q^2a^2t^2 + \Delta^2\right)^{3/2}} \left[\left(\frac{1}{27^3}\right) \sqrt{\frac{3}{2}}\frac{1}{2\pi a_0^2}\int x'^5   \exp\left(\frac{-2x'\sqrt{1+z'^2}}{3}\right)\left(-1+2z'^2\right)dz'\right] \label{eq:final}
\end{eqnarray}
The integral in Eq. \ref{eq:final} must be evaluated numerically for each value of $x'$.  The resulting dimensionless form of the function in brackets is shown in Fig. \ref{fig:delta_rho}.
\begin{figure}
    \includegraphics[width=.5\columnwidth]{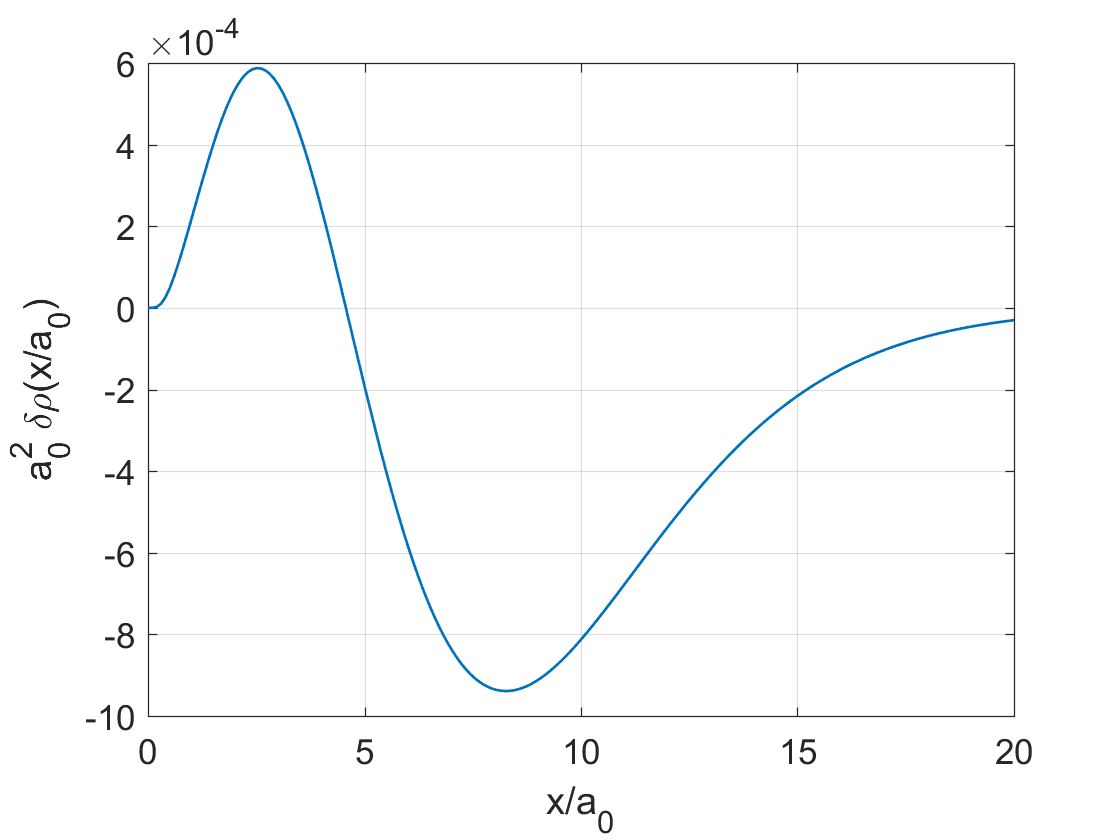}
	\caption{Plot of term in brackets of Eq. \ref{eq:final}.  The position of the maximum of the absolute value of this curve can be used to estimate the effective Bohr radius $a_0$, and the max value of the curve determines the magnitude of the coupling coefficient between $d_{z^2}$ and $d_{x^2-y^2}$ orbitals.  This in turn is related to the magnitude of the Berry curvature.}\label{fig:delta_rho}
\end{figure}
The position of the minimum is at $x\approx 8.3 a_0$, and the value of the function there (in dimension-ful form) is $-9.39\times 10^{-4}\times \left(\frac{1}{a_0}\right)^2$.  We combine these two facts to obtain:
\begin{eqnarray}
\int dz ~\frac{\delta \rho(q,x_{\rm min},0,z)}{eE} &=&  -\frac{ 2 \Delta a t }{\left(4q^2a^2t^2 + \Delta^2\right)^{3/2}} ~0.065\times\left(\frac{1}{x_{\rm min}^2}\right) \label{eq:rho2d}
\end{eqnarray}
Using the expression for $\Omega(q)$ in Eq.~\ref{eq:Omegaq}, we arrive at Eq.~14 of the main text:
\begin{eqnarray}
-\int dz~\frac{\delta \rho(q,x_{\rm min},0,z)}{eE} \times \frac{a t x_{\rm min}^2}{0.065} &=&  \nu~ \Omega_\nu(q)
\end{eqnarray}

\subsection{Energy-resolved Hall conductivity}

We next discuss the energy-resolved Berry curvature, and the energy-resolved change in charge density distribution.  We will find a simple relationship between these two quantities.
The general expression for the energy-resolved valley Hall and orbital Hall conductivity are:
\begin{eqnarray}
\sigma_{\rm VHE}(\varepsilon) &=& 2~ {\rm Im}\int d{\bf k} \sum_{n,m} (f_{n,{\bf k}}-f_{m,{\bf k}}) ~\tau_{\bf k} ~\frac{v_{nm}^x v_{mn}^y}{\left(E_n - E_m\right)^2}~ \delta\left(\varepsilon - E_{n,\bf k}\right) \label{eq:VHE_E}\\
\sigma_{Lz}(\varepsilon) &=& 2~ {\rm Im}\int d{\bf k} \sum_{n,m} (f_{n,{\bf k}}-f_{m,{\bf k}}) ~\frac{v_{nm}^x v_{mn}^{Lz}}{\left(E_n - E_m\right)^2} ~\delta\left(\varepsilon - E_{n,\bf k}\right) \label{eq:OHE_E}
\end{eqnarray}
where $v_{nm}^x=\langle \psi_n | \hat{v}_x | \psi_m \rangle$, $v_{nm}^{Lz}=\langle \psi_n | \left(L_z \hat{v}_y + \hat{v}_y Lz\right)/2 | \psi_m \rangle$, and $\hat{v}_x = dH/dk_x$.  In Eq. \ref{eq:VHE_E}, $\tau_{\bf k}$ is $\pm 1$ according to the state's valley index.\\

\begin{figure}
	\includegraphics[width=.9\columnwidth]{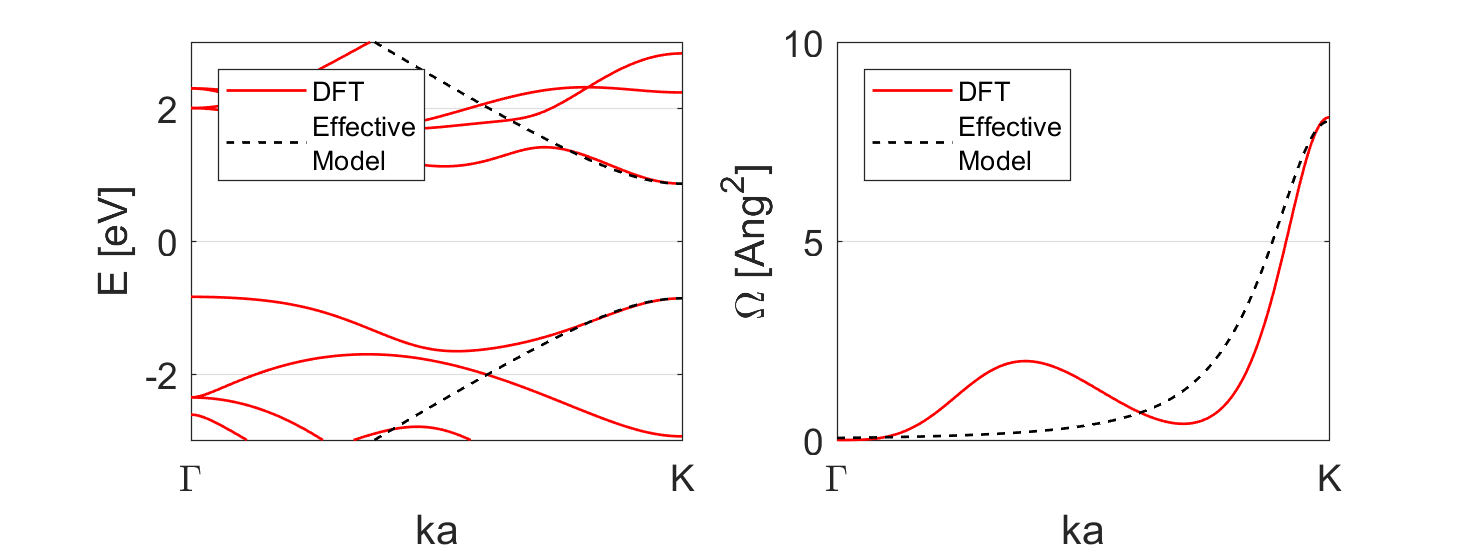}
	\caption{Comparison of band structure and Berry curvature of valence band computed with density functional theory, and with the effective model.  The parameters of the effective model are chosen to reproduce the band dispersion near ${\bf K}$, and are $t=1.09~{\rm eV}$, $\Delta=1.7~{\rm eV}$.}\label{fig:delta_rho}
\end{figure}

Using Eq. \ref{eq:rho2d}, we obtain:
\begin{eqnarray}
\frac{\delta \rho(q,x_{\rm min})}{eE} \times \frac{a t x_{\rm min}^2}{0.065}=  -\frac{ 2 \Delta (at)^2 }{\left(4q^2a^2t^2 + \Delta^2\right)^{3/2}} \label{eq:deltarhoq}
\end{eqnarray}

Next we make the connection between the real space density and the Berry curvature.  To do so, we note that the expression on the right-hand-side of Eq.~\ref{eq:deltarhoq} is equal to the $q$-dependent valley Hall conductivity, Eq.~\ref{eq:Omegaq}:
\begin{eqnarray}
\sigma_{\rm VHE}(\varepsilon)&=& -\frac{a t x_{\rm min}^2}{0.065} \times \frac{2}{(2 \pi)}\int dz~ \int q~ d q ~\frac{\delta \rho(q,x_{\rm min},0,z)}{eE} ~ \delta\left(\varepsilon-\frac{1}{2}\sqrt{4q^{2}a^2 t^{2} +\Delta^2}\right)
\end{eqnarray}
So that:
\begin{eqnarray}
\sigma_{\rm VHE}(\varepsilon)&=&-\frac{at x_{\rm min}^2}{0.065} \left( \frac{\delta {\rm LDOS}(\varepsilon,x_{\rm min})}{eE} \right)
\end{eqnarray}


 \section{Calibration of the local density of states measurement}

\begin{figure}
 	\includegraphics[width=.9\columnwidth]{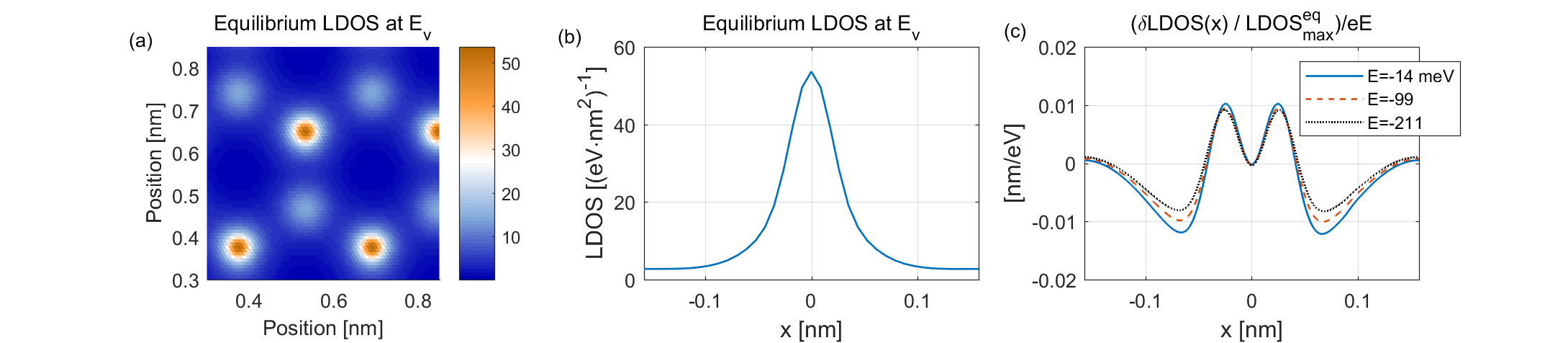}
 	\caption{(a) Equilibrium local density of states at the valence band edge energy. (b) 1-dimensional cut through the local density of states through the maxmimum, which is located on the Mo atom. (c) is a replot of the data from Fig. 4(d) of the main text, normalized by the maximum value of the equilibrium density of states.}\label{fig:LDOSeq}
 \end{figure}

The differential conductance measured in an STM experiment is proportional to the local density of states, and the constant of proportionality may depend on experimental details and may be unknown.  However, Eq. (13) of the main text relies on the absolute value of the local density of states.  In order to calibrate the measurement, we provide the equilibrium local density of states at the valence band edge energy in Fig. \ref{fig:LDOSeq}.  As before, we assume that the measured signal is proportional to the integral over the $z$ coordinate.  The local density of states exhibits a maximum value at the center of the Mo atom, with a value of $53~{\rm (eV\cdot nm^2)^{-1}}$.  Fig. \ref{fig:LDOSeq}(c) shows the electric field induced change in the density of states, normalized by this maximum value.  By normalizing the data this way, the unknown constant of proportionality between the signal and the local density of states factors out of the data.

The unit of the normalized change in LDOS is inverse electric field, and provides the fractional change of the local density of states for a given applied electric field.  To achieve a fractional change of $10^{-4}$ requires an applied field of $10^{2}~{\rm V/nm}$.  As mentioned in the main text, this is on the order of the breakdown field measured for MoS$_2$ \cite{lembke2012breakdown}, so that the effect we describe is quantifiable for measurements with signal to noise ratio of less than $10^{-4}$.

\end{document}